# Modulation of spin conversion in a 1.5 nm-thick Pd film by ionic gating

Shin-ichiro Yoshitake, Ryo Ohshima, Teruya Shinjo, Yuichiro Ando, and Masashi Shiraishi

Department of Electronic Science and Engineering, Kyoto University, Japan.

# Corresponding author: Masashi Shiraishi (shiraishi.masashi.4w@kyoto-u.ac.jp)

**Abstract**

Gate-induced modulation of the spin–orbit interaction (SOI) in a 1.5 nm-thick Pd thin film grown on a ferrimagnetic insulator was investigated. Efficient charge accumulation by ionic gating enables a substantial upshift in the Fermi level of the Pd film, which was corroborated by suppression of the resistivity in the Pd. Electromotive forces arising from the inverse spin Hall effect in Pd under spin pumping were substantially modulated by the gating, in consequence of the modulation of the spin Hall conductivity of Pd as in an ultrathin Pt film. The same experiment using a thin Cu film, for which the band structure is largely different from Pd and Pt and its SOI is quite small, provides further results supporting our claim. The results obtained help in developing a holistic understanding of the gate-tunable SOI in solids and confirm a previous explanation of the significant modulation of the spin Hall conductivity in an ultrathin Pt film by gating.

Electric gating using a solid gate insulator is the most pivotal technical tool in modern electronics. An efficient modulation of the conductivity of a channel layer of a field-effect transistor (FET) by solid gating is the key operating principle of FETs. Gating by ionic liquids or gels, which is a modern advanced technique for electric gating, has been garnering attention in a broad range of condensed-matter physics fields. Compared with solid gating, ionic gating enables more efficient charge accumulation in adjacent solids, which in turn enables pioneering explorations that have led to discoveries such as the appearance of a superconducting state in an oxide insulator [1] and in transition-metal dichalcogenides [2], the paramagnetic–ferromagnetic transition in InMsAs [3], strong modulation of the Curie temperature of nanometer-thick Co (ultrathin Co) [4], and the vanishing of the inverse spin Hall effect (ISHE) through the large modulation of the spin-orbit interaction (SOI) and the spin Hall conductivity in ultrathin Pt [5]. The formation of a thin electric double layer [6] facilitated these notable achievements.

Among these achievements, a combination of ultrathin Pt films and ionic gating opened a new frontier in spin conversion science, which had been gathering tremendous attention because spin conversion [7] allowed the development of, for example, spin-torque diodes [8] and spin-orbit torque switching devices [9,10]. Tunable spin conversion expands this research field. Both resistance and the spin Hall conductivity in a 2 nm-thick Pt film, in which the intrinsic spin Hall effect governs the spin conversion, were simultaneously modulated by ionic gating [5]. This achievement was attributed to a sufficient charge accumulation in the Pt, resulting in a large upshift of the Fermi level by the ionic gating. The ISHE is governed by the density of states (DOS) of the $d$-orbitals [11,12], and the DOS of the $d$-orbitals of Pt vanishes above the Fermi level [13]. This physics is responsible for the suppression of the spin Hall conductivity, i.e., the vanishing of the ISHE by positive gating. The formation of an ultrathin film with a low charge carrier density and the use of ionic gating for efficient charge accumulation are key to the tunable ISHE.

Given this brief survey of studies on spin conversion and its control by gating, acquiring additional insights and more detailed understanding of the physics in spin conversion science using gating effects by combining ultrathin Pd films and ionic gating can represent a substantial contribution to the spintronics field because the band structure of Pd is quite similar to Pt [13]. Indeed, a similar gate-tunable SOI can appear if the current understanding of the physics of gate-tunable SOI is correct. In the present study, thin Pd films were prepared and subsequently equipped with ionic gates. A similar modulation of the resistance and an electromotive force (EMF) due to spin conversion was detected in the Pd using the ionic gating technique, which underscores validity of the understanding of the phenomenon constructed previously [5]. To support our assertion, the same experiments using thin Cu films were also implemented with no suppression of the EMF for the gating.

Figure 1(a) shows a schematic of the fabricated sample and the measurement setup. Yttrium–iron–garnet (YIG) grown on 0.7 μm-thick, 3 mm-long, and 1 mm-wide gadolinium–gallium–garnet (GGG) (Granopt, Japan) was used as a spin source. The YIG was polished with an agglomerate-free alumina polishing suspension (50 nm particle size) for 40 min and then annealed at 1273 K in air for 90 min. The Pd was deposited onto the YIG at a rate of 0.02 nm/s via electron-beam evaporation. The thickness of the Pd film was changed from 1.5 to 30 nm. Afterwards, Ti(5 nm)/Au(50 nm) electric pads were formed on the sides of the sample via electron-beam evaporation. An ionic gel was prepared by mixing PS–PMMA–PS polymer (Polymer Source, USA), DEME–TFSI ionic liquid (Kanto Chemical, Japan), and ethyl propionate ($CH_3CH_2COOC_2H_5$, Nacalai Tesque, Japan) in a weight ratio of 9.3:0.7:20, respectively. Insulating double-sided adhesive tape was placed on the sides of the Pd channel (inside the area covered by Ti/Au electric pads) to provide additional mechanical support for the gate electrode film on top of which it was placed. The gate electrode film was mounted directly above the Pd channel after the ionic gel was applied.

For the measurement, a sample was mounted in a $TE_{011}$ cavity of an electron-spin resonance system (JEOL JES-FA200). The applied microwave power was set to 10 mW; and the microwave frequency was 9.12 GHz. The gate voltage was applied at room temperature; after the ionic gel developed an electric double layer, the sample was cooled to 250 K and resistivity, ferromagnetic resonance (FMR), and EMF measurements were carried out after the gate leakage current was confirmed to be suppressed; i.e., ions in the ionic gel were not mobile. The sample temperature was then increased to room temperature, the gate voltage was changed, and the sample was again cooled to 250 K. A flashing flow of $N_2$ gas was supplied to the cavity containing the sample. EMFs due to spin conversion in the Pd were monitored under an external dc magnetic field along 0° and 180°, and the EMF at 180° was subtracted from that at 0° to eliminate unwanted thermal contributions and determine the precise amplitude of the EMF.

Figure 1(b) shows the dependence of the resistivity of the Pd films; the resistivity monotonically increases with decreasing film thickness. Thickness dependence of the resistivity in thin metallic films can be reproduced by a theoretical fitting function [14],

$$\rho = \rho_{bulk}[1 - \left(\frac{1}{2} + \frac{3}{4}\frac{\lambda}{t}\right)(1 - pe^{-\xi t/\lambda})e^{-t/\lambda}]^{-1}, \quad (1)$$

where $\rho$, $\rho_{bulk}$, $\lambda$, $t$, $\xi$, and $p$ are the resistivity of the film, the bulk resistivity of Pd (22 μΩ cm in the fitting result), the electron mean free path, the sample thickness, the grain-boundary penetration parameter, and the fraction of carriers specularly scattered at the surface, respectively. As shown in Fig. 1(b), the experimental result is well reproduced using this fitting function. From the best fit using Eq. (1), $\lambda$, $\xi$, and $p$ were estimated to be 44 nm, 0.035, and 0.97, respectively. From the fitting result, the increase in resistivity of the thinner Pd films is rationalized by enhancement of surface scattering, as reported for Pt [5] and other metals [14]. More importantly, the 1.5 nm-thick Pd film is continuous because the theoretical fitting holds in the thinnest case, its resistivity being 1250 μΩ cm, which indicates that the 1.5 nm-thick Pd film operates in the intrinsic spin Hall regime at room temperature [15] as designed. From here on, we focus on the 1.5 nm-thick Pd because gating effects are more salient in thinner films as explained above.

Figure 2(a) shows the gate dependence of the resistance of the 1.5 nm-thick Pd film. Although a negative gate voltage can be applied, we shed light mainly on the results under an applied positive gate voltage because the Pd film degrades under negative gate voltage applications (see Supplemental Materials) due to unwanted surface reactions with water molecules present in the ionic liquid [6]. The resistance decreases from roughly 3.3 kΩ to 2.0 kΩ, the fraction decrease being 40%. The decrease in resistivity is attributed to efficient charge accumulation as in ultrathin Pt films [5]. Figure 3 shows the FMR spectra and EMF from the 1.5 nm-thick Pd at external magnetic fields of 0° and 180° as a function of gate voltages (see also Supplemental Materials for line-shape analyses for the FMR spectra and the EMFs). Prominent EMFs are measured at 0° and 180° under FMR of the YIG; more importantly, their polarities are opposite to each other, which is a compelling result that the measured EMFs are ascribed to the ISHE for Pd. The amplitude of the EMF at $V_g = 0$ V is estimated to be 3 μV, which is comparable to that observed for ultrathin Pt (2 μV for the 2 nm-thick Pt [5]). Furthermore, the amplitude of the EMFs monotonically decreases for positive gate voltages (see also Fig. 2(b)). The sudden decrease in the EMF at $V_g = 1.25$ V is ascribable to the waiting time of the measurements, which were long because we needed to refill the liquid $N_2$.

      The dependences of the electric current in the Pd film and the normalized electric current on gate voltage are shown in Figs. 2(c) and 2(d), where normalization was applied using the amplitude at $V_g = 0$ V. Because the ISHE is an effect where electric current is generated by the conversion from a spin current induced by the SOI, and not an effect for which voltage is generated, the EMFs were converted to an electric current by considering the resistivity modulation as a function of the gate voltage. As is evident, the electric current monotonically decreases for positive gate voltages and the suppression ratio of the electric current is about 80%. Given that the band structure of Pd is quite similar with that of Pt, the results in the study are understood as follows: the intrinsic mechanism governs the spin conversion in the 1.5 nm-thick Pd within the spin-Hall-based spin conversion regime [16]. The intrinsic spin Hall effect (and its reciprocal effect, ISHE) originates from the inter-$d$-band excitation, as previously described. Because the ISHE observed in an ultrathin Pt film is also intrinsic in nature, the shift of the Fermi level to a position where the DOS of the $d$-orbitals is largely suppressed is a key factor

in the suppression of the ISHE due to that of the spin Hall conductivity [5]. The same mechanism holds in the 1.5 nm-thick Pd; i.e., these results provide additional compelling evidence for a gate-tunable SOI realized in an ultrathin Pd film as for an ultrathin Pt film.

For further supporting evidence, we performed the same experiment using a thin Cu film. Unlike Pt and Pd, Cu possesses a weak and constant *d*-orbital contribution at and above the Fermi level [16-18], resulting in a small and non-tunable spin Hall conductivity. Therefore, an ISHE-induced electric current from a thin Cu film is thought to be quite weak and almost non-tunable. In the experiment, an 8.1 nm-thick Cu film was used because surface oxidation takes place quickly (see also Supplemental Materials for details of the experiments). Results concerning the gate voltage dependence of the resistance and normalized electric current generated by the ISHE are shown in Figs. 4(a) and 4(b). The resistance of the Cu film monotonically decreases albeit weakly for the positive gate voltages as observed in the Pd (this study) and the Pt [5]. A weak gate voltage dependence of the resistance of the Cu occurs because the Cu film is considerably thicker than those of the Pd and the Pt [5]. With the modulation of the resistance being realized as planned, an upshift in the Fermi level was expected given an efficient carrier accumulation in the Cu film. However, the ISHE-induced electric current does not exhibit a salient gate-voltage dependence because of its band structure. It has been well known that surface oxidation of Cu and formation of CuO/Cu give rise to sizable spin-to-charge conversion even in Cu [19] and an application of a strong gate voltage enables $O^{2-}$ migration to the top surface of Pt [20], resulting in enhancement of spin-to-charge conversion. Given that (i) ionic gating allows more efficient generation of an electric field, probably enhancing $O^{2-}$ migration to the top surface of the Cu in our study, (ii) the surface of the Cu in this study is most likely naturally oxidized and (iii) thicknesses of the Cu and the Pt were comparable to that of the Cu in our study, it is intriguing that such enhancement was not observed and the gate dependence of the ISHE was as expected for Cu. Further study is still awaited. Anyhow, the difference in the ISHE-induced electric current for gate voltages applied to the Pd and the Cu thin films validates our holistic understandings of the gate-tunable SOI for ultrathin single metal.

Finally, we briefly discuss plausible applications of the gate-tunable ISHE observed in Pd and Pt based on the holistic understandings established in this study. An example is an application for spin-orbit torque magnetic random-access memories (SOT-MRAMs). SOT-MRAMs possesses superiority in endurance comparing to conventional MRAMs because spin current gives much less damages to tunneling barriers in magnetic tunnel junctions (MTJs) and the spin current is generated by using the SOI of heavy materials. In a conventional SOT-MRAMs, an MTJ is aligned at the crossing point of a word line and a bit line to control spin current injection. Since the SOI is gate-tunable as we have been clarifying, we need only one line to store and rewrite information by equipping a gate electrode to control the SOI by the gating. Furthermore, since the SOI modulation allows modulation of spin diffusion length, the gate-tunable SOI may allow creation of a spin transistor using metallic materials. In addition, further investigation for a quest of appropriate material selections and combinations are awaited, because energy dependence SOI is theoretically predicted in Fe- and Pt-doped Au [21] and they can be the other potential candidate for generating the gate-tunable ISHE.

In summary, the gate-induced modulation of the SOI in 1.5 nm-thick Pd thin film grown on a ferrimagnetic insulator was investigated. Efficient charge accumulation due to ionic gating enabled a substantial upshift in the Fermi level of the Pd films, allowing substantial suppression of the EMFs arising from the ISHE in Pd because of the modulation of the spin Hall conductivity by the gating. The same experiment using a thin Cu film revealed no sizable modulation of the EMFs in Cu by gating. Considering the band structure and the small SOI of Cu, the weak gating effect of the EMFs from the Cu is rationalized by the physics of a gate-tunable SOI observed in Pd and Pt. Hence, the whole results described in the paper support a holistic understanding of the gate-tunable SOI in solids and corroborates previous explanations of the significant modulation of the spin Hall conductivity in an ultrathin Pt films by gating [5].

Supplemental Material describes the surface reaction of the thin Pd under an application of negative gate voltage, details of the experiments using thin Cu films, and line-shape analyses of the

FMR spectra from the YIG and EMFs from the Pd and the Cu.

The authors acknowledge the support by a Grant-in-Aid for Scientific Research (S) No. 16H06330, "Semiconductor spincurrentronics", and MEXT (Innovative Area "Nano Spin Conversion Science" KAKENHI No. 26103003).

The data that support the findings of this study are available from the corresponding author upon reasonable request.

**Figures and Figure captions**

**Figures**

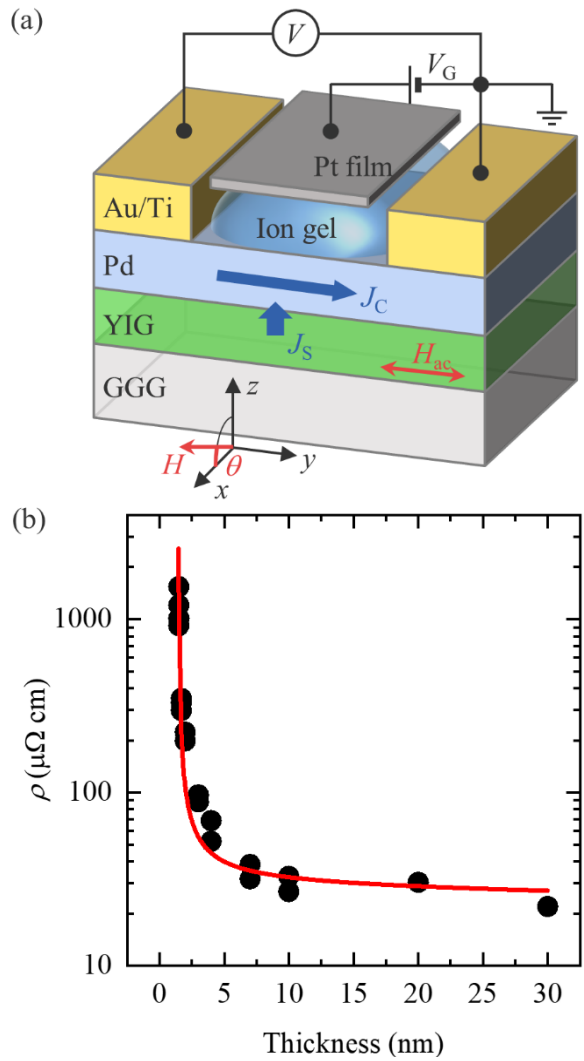

Fig. 1 Yoshitake et al.

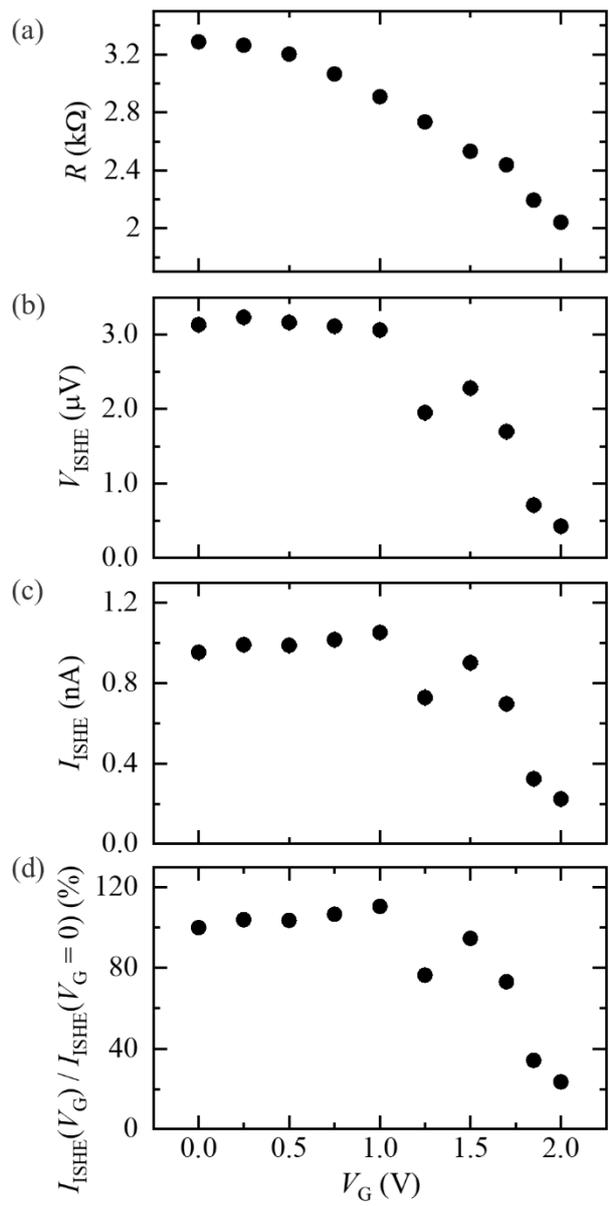

Fig. 2 Yoshitake et al.

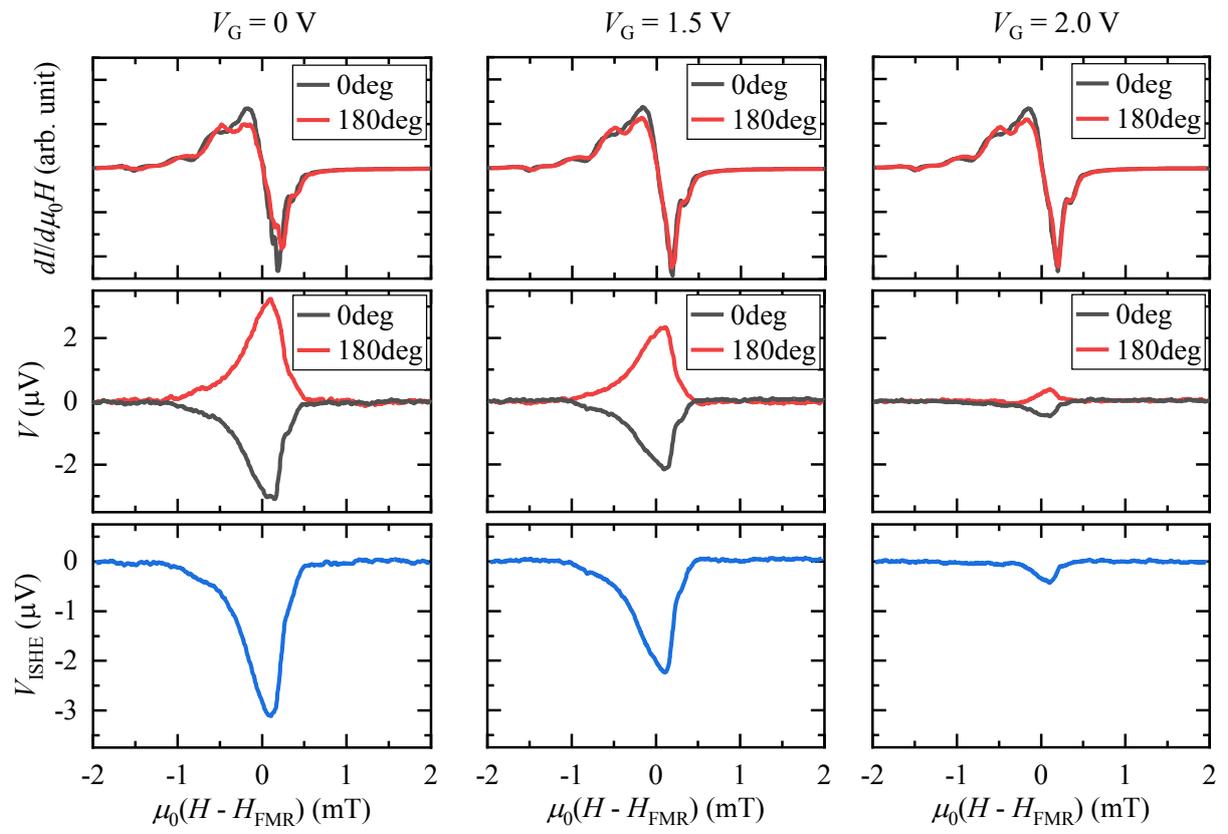

Fig. 3 Yoshitake et al.

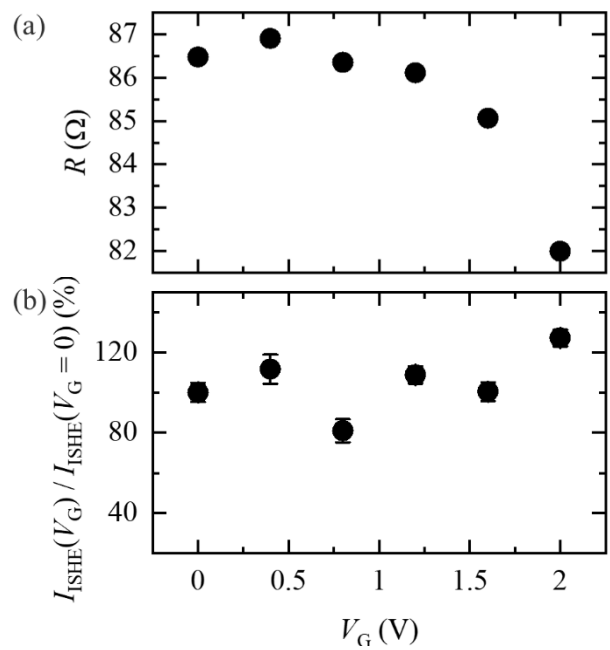

Fig. 4 Yoshitake et al.

**Figure captions**

**Fig. 1 (a)** Schematic of ultrathin Pd sample for gate-tunable spin conversion measurements. **(b)** Thickness dependence of the resistivity of Pd thin films. The red solid line is the result of the theoretical fitting using Eq. (1) in the main text.

**Fig. 2** Gate voltage dependences of **(a)** resistances, **(b)** EMFs, **(c)** magnitudes of the electric current generated by the ISHE, and **(d)** the normalized electric current, of the 1.5 nm-thick Pd.

**Fig. 3** Observed FMR spectra and EMFs of the 1.5 nm-thick Pd under the FMR of the YIG, when the gate voltage is changed from 0 to +2 V in increments of +1 V. The upper panels show FMR spectra of the YIG, the middle panels show raw data when the external magnetic field was applied at 0° (the black solid line) and 180° (the red solid lie), and the lower panels show the net EMFs for each gate voltage. See also the main text how to estimate the net EMFs.

**Fig. 4** Gate voltage dependences of **(a)** resistances, and **(b)** the normalized electric current, of the 8.1 nm-thick Cu.